\documentclass[aps, prb, reprint, superscriptaddress]{revtex4-1}
\usepackage[english]{babel}
\usepackage{amsmath,amsthm}
\usepackage{amsfonts}
\usepackage{xspace}
\usepackage{bm}
\usepackage{booktabs}
\usepackage{graphicx}
\usepackage{dcolumn}
\usepackage[mathlines]{lineno}
\usepackage[colorlinks]{hyperref}  
\usepackage{gensymb}
\usepackage{epstopdf}
\usepackage{newfloat}
\usepackage[flushleft]{threeparttable}
\usepackage{color}

\DeclareFloatingEnvironment[name={Figure S}]{suppfigure}
\DeclareFloatingEnvironment[name={S}]{suppEquation}

\newcommand*{\citen}[1]{%
  \begingroup
    \romannumeral-`\x 
    \setcitestyle{numbers}%
    \cite{#1}%
  \endgroup   
}

\newcommand{\muB}{\ensuremath{\mu_\mathrm{B}}\xspace}
\newcommand{\HFMR}{\ensuremath{H_\mathrm{FMR}}\xspace}
\newcommand{\Hcub}{\ensuremath{H_\mathrm{||,4}}\xspace}
\newcommand{\DelH}{\ensuremath{\Delta H}\xspace}
\newcommand{\DelHtm}{\ensuremath{\Delta H_\mathrm{2m}}\xspace}
\newcommand{\DelHo}{\ensuremath{\Delta H_\mathrm{0}}\xspace}

\newcommand{\DelHtmhkl}{\ensuremath{\Delta H^\mathrm{\langle hk0\rangle}_\mathrm{2m}}\xspace}
\newcommand{\Meff}{\ensuremath{M_\mathrm{eff}}\xspace}
\newcommand{\Ms}{\ensuremath{M_\mathrm{s}}\xspace}
\newcommand{\gOP}{\ensuremath{g_\mathrm{op}}\xspace}
\newcommand{\gIP}{\ensuremath{g_\mathrm{ip}}\xspace}
\newcommand{\tSRO}{\ensuremath{t_\mathrm{SRO}}\xspace}
\newcommand{\tTh}{\ensuremath{t_\mathrm{th}}\xspace}
\newcommand{\rSRO}{\ensuremath{\rho_\mathrm{SRO}}\xspace}
\newcommand{\rbulk}{\ensuremath{\rho_\mathrm{b}}\xspace}
\newcommand{\rsurf}{\ensuremath{\rho_\mathrm{s}}\xspace}
\newcommand{\tLSMO}{\ensuremath{t_\mathrm{LSMO}}\xspace}
\newcommand{\Gmix}{\ensuremath{G_\mathrm{\uparrow\downarrow}}\xspace}
\newcommand{\Gext}{\ensuremath{G_\mathrm{ext}}\xspace}
\newcommand{\sdl}{\ensuremath{\lambda_\mathrm{s}}\xspace}

\newcommand{\Gtm}{\ensuremath{\Gamma_\mathrm{2m}}\xspace}
\newcommand{\fM}{\ensuremath{f_\mathrm{M}}\xspace}

\begin{document}

\title{Spin transport and dynamics in all-oxide perovskite La$_{2/3}$Sr$_{1/3}$MnO$_3$/SrRuO$_3$ bilayers probed by ferromagnetic resonance}%

\author{Satoru Emori}%
\email{
satorue@stanford.edu
}
\affiliation{ 
Geballe Laboratory for Advanced Materials, Stanford University, Stanford, CA 94305 USA
}%
\author{Urusa S. Alaan}
\affiliation{ 
Geballe Laboratory for Advanced Materials, Stanford University, Stanford, CA 94305 USA
}%
\affiliation{ 
Department of Materials Science and Engineering, Stanford University, Stanford, CA 94305 USA
}%
\author{Matthew T. Gray}
\affiliation{ 
Geballe Laboratory for Advanced Materials, Stanford University, Stanford, CA 94305 USA
}%
\affiliation{ 
Department of Materials Science and Engineering, Stanford University, Stanford, CA 94305 USA
}%
\author{Volker Sluka}
\affiliation{ 
Department of Physics, New York University, New York, NY 10003, USA
}%
\author{Yizhang Chen}
\affiliation{ 
Department of Physics, New York University, New York, NY 10003, USA
}%
\author{Andrew D. Kent}
\affiliation{ 
Department of Physics, New York University, New York, NY 10003, USA
}%
\author{Yuri Suzuki}
\affiliation{ 
Geballe Laboratory for Advanced Materials, Stanford University, Stanford, CA 94305 USA
}%
\affiliation{ 
Department of Applied Physics, Stanford University, Stanford, CA 94305 USA
}%

\date{\today}

\begin{abstract}
Thin films of perovskite oxides offer the possibility of combining emerging concepts of strongly correlated electron phenomena and spin current in magnetic devices. 
However, spin transport and magnetization dynamics in these complex oxide materials are not well understood. 
Here, we experimentally quantify spin transport parameters and magnetization damping in epitaxial perovskite ferromagnet/paramagnet bilayers of La$_{2/3}$Sr$_{1/3}$MnO$_3$/SrRuO$_3$ (LSMO/SRO) by broadband ferromagnetic resonance spectroscopy. 
From the SRO thickness dependence of Gilbert damping, we estimate a short spin diffusion length of $\lesssim$1 nm in SRO and an interfacial spin-mixing conductance comparable to other ferromagnet/paramagnetic-metal bilayers.  
Moreover, we find that anisotropic non-Gilbert damping due to two-magnon scattering also increases with the addition of SRO. 
Our results demonstrate LSMO/SRO as a spin-source/spin-sink system that may be a foundation for examining spin-current transport in various perovskite heterostructures.

\end{abstract}
\maketitle

\section{Introduction}
Manipulation and transmission of information by spin current is a promising route toward energy-efficient memory and computation devices~\cite{Stamps2014}. 
Such spintronic devices may consist of ferromagnets interfaced with nonmagnetic conductors that exhibit spin-Hall and related spin-orbit effects~\cite{Hoffmann2013, Sinova2015, Hoffmann2015a}.
The direct spin-Hall effect in the conductor can convert a charge current to a spin current, which exerts torques on the adjacent magnetization and modifies the state of the device~\cite{Brataas2012, Locatelli2014}. 
Conversely, the inverse spin-Hall effect in the conductor can convert a propagating spin current in the magnetic medium to an electric signal to read spin-based information packets~\cite{Chumak2015}.  
For these device schemes, it is essential to understand the transmission of spin current between the ferromagnet and the conductor, which is parameterized by the spin-mixing conductance and spin diffusion length. 
These spin transport parameters can be estimated by spin pumping at ferromagnetic resonance (FMR), in which a spin current is resonantly generated in the ferromagnet and absorbed in the adjacent conductor~\cite{Tserkovnyak2002a, Tserkovnyak2002}. 
Spin pumping has been demonstrated in various combinations of materials, where the magnetic layer may be an alloy (e.g., permalloy) or insulator (e.g., yttrium iron garnet) and the nonmagnetic conductor may be a transition metal, semiconductor, conductive polymer, or topological insulator~\cite{Czeschka2011, Sun2013, Zhang2015a, Du2015, Ando2011a, Ando2013, Jamali2015}.  

Transition metal oxides, particularly those with the perovskite structure, offer the intriguing prospect of integrating a wide variety of strongly correlated electron phenomena~\cite{Zubko2011, Hwang2012} with spintronic functionalities~\cite{Majumdar2014,Lesne2016}.
Among these complex oxides, La$_{2/3}$Sr$_{1/3}$MnO$_3$ (LSMO) and SrRuO$_3$ (SRO) are attractive materials for epitaxial, lattice-matched spin-source/spin-sink heterostructures. 
LSMO, a metallic ferromagnet known for its colossal magnetoresistance and Curie temperature of $>$300 K, can be an excellent resonantly-excited spin source because of its low magnetization damping~\cite{Luo2013, Luo2015, Lee2016, Haidar2015, Atsarkin2016, Wahler2016}. 
SRO, a room-temperature metallic paramagnet with relatively high conductivity~\cite{Koster2012}, exhibits strong spin-orbit coupling~\cite{Langner2009} that may be useful for emerging spintronic applications that leverage spin-orbit effects~\cite{Hoffmann2013, Sinova2015, Hoffmann2015a}.

A few recent studies have reported dc voltages at FMR in LSMO/SRO bilayers that are attributed to the inverse spin-Hall effect in SRO generated by spin pumping~\cite{Haidar2015, Atsarkin2016, Wahler2016}. 
However, it is generally a challenge to separate the inverse spin-Hall signal from the spin rectification signal, which is caused by an oscillating magnetoresistance mixing with a microwave current in the conductive magnetic layer~\cite{Azevedo2011, Bai2013, Obstbaum2014}. 
Moreover, while the spin-mixing conductance is typically estimated from the enhancement in the Gilbert damping parameter $\alpha$, the quantification of $\alpha$ is not necessarily straightforward in epitaxial thin films that exhibit pronounced anisotropic non-Gilbert damping~\cite{Lindner2003a, Woltersdorf2004a,Lenz2006, Zakeri2007a, Barsukov2011, Kurebayashi2013,Lee2016}. 
It has also been unclear how the Gilbert and non-Gilbert components of damping in LSMO are each modified by an adjacent SRO layer.
These points above highlight the need for an alternative experimental approach for characterizing spin transport and magnetization dynamics in LSMO/SRO.

In this work, we quantify spin transport parameters and magnetization damping in epitaxial LSMO/SRO bilayers by broadband FMR spectroscopy with \textit{out-of-plane} and \textit{in-plane} external magnetic fields. 
Out-of-plane FMR enables straightforward extraction of Gilbert damping as a function of SRO overlayer thickness, which is reproduced by a simple ``spin circuit'' model based on diffusive spin transport~\cite{Boone2013, Boone2015}. 
We find that the spin-mixing conductance at the LSMO/SRO interface is comparable to other ferromagnet/conductor interfaces and that spin current is absorbed within a short length scale of $\lesssim$1 nm in the conductive SRO layer. 
From in-plane FMR, we observe pronounced non-Gilbert damping that is anisotropic and scales nonlinearly with excitation frequency,  which is accounted for by an existing model of two-magnon scattering~\cite{Arias1999}.  
This two-magnon scattering is also enhanced with the addition of the SRO overlayer possibly due to spin pumping. 
Our findings reveal key features of spin dynamics and transport in the prototypical perovskite ferromagnet/conductor bilayer of LSMO/SRO and provide a foundation for future all-oxide spintronic devices.

\section{Sample and Experimental Details}

Epitaxial films of LSMO(/SRO) were grown on as-received (001)-oriented single-crystal (LaAlO$_3$)$_{0.3}$ (Sr$_2$AlTaO$_6$)$_{0.7}$ (LSAT) substrates using pulsed laser deposition. 
LSAT exhibits a lower dielectric constant than the commonly used SrTiO$_3$ substrate and is therefore better suited for high-frequency FMR measurements.  
The lattice parameter of LSAT (3.87 \AA) is also closely matched to the pseudocubic lattice parameter of LSMO ($\approx$3.88 \AA).
By using deposition parameters similar to those in previous studies from our group~\cite{Takamura2008, Grutter2010}, all films were deposited at a substrate temperature of 750 $^\circ$C with a target-to-substrate separation of 75 mm, laser fluence of $\approx$1 J/cm$^2$, and repetition rate of 1 Hz. 
LSMO was deposited in 320 mTorr O$_2$, followed by SRO in 100 mTorr O$_2$. 
After deposition, the samples were held at 600  $^\circ$C for 15 minutes in $\approx$150 Torr O$_2$ and then the substrate heater was switched off to cool to room temperature. The deposition rates were calibrated by x-ray reflectivity measurements. 
The thickness of LSMO, \tLSMO, in this study is fixed at 10 nm, which is close to the minimum thickness at which the near-bulk saturation magnetization can be attained.    

\begin{figure}[tb]
  \includegraphics [width=0.97\columnwidth] {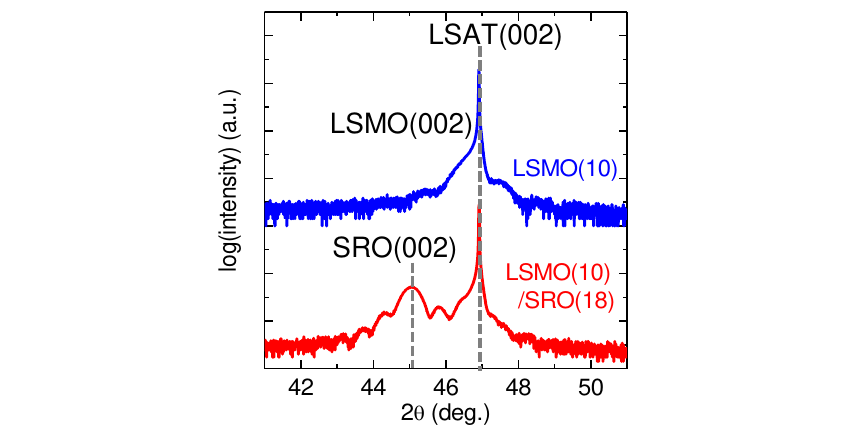}
  \centering
  \caption{\label{fig:XRD}
    2$\theta$-$\omega$ x-ray diffraction scans of a single-layer LSMO(10 nm) film and LSMO(10 nm)/SRO(18 nm) bilayer.}
\end{figure}

X-ray diffraction results indicate that both the LSMO films and LSMO/SRO bilayers are highly crystalline and epitaxial with the LSAT(001) substrate, with high-resolution 2$\theta$-$\omega$ scans showing distinct Laue fringes around the (002) Bragg reflection (Fig. 1).
In this study, the maximum thickness of the LSMO and SRO layers combined is less than 30 nm and below the threshold thickness for the onset of structural relaxation by misfit dislocation formation~\cite{Takamura2008,Grutter2010}.  
The typical surface roughness of LSMO and SRO measured by atomic force microscopy is $\lesssim$4 \AA, comparable to the roughness of the LSAT substrate surface. 

SQUID magnetometry confirms that the Curie temperature of the LSMO layer is $\approx$350 K and the room-temperature saturation magnetization is $\Ms\approx300$ kA/m for 10-nm thick LSMO films.  
The small LSMO thickness is desirable for maximizing the spin-pumping-induced enhancement in damping, since spin pumping scales inversely with the ferromagnetic layer thickness~\cite{Tserkovnyak2002a, Tserkovnyak2002}. 
Moreover, the thickness of 10 nm is within a factor of $\approx$2 of the characteristic exchange length $\sqrt{2A_\mathrm{ex}/\mu_0 \Ms^2}\approx5$ nm, assuming an exchange constant of $A_\mathrm{ex}\approx2$ pJ/m in LSMO (Ref.~\citen{Golosovsky2007}), so standing spin-wave modes are not expected. 

\begin{figure}[tb]
  \includegraphics [width=1.00\columnwidth] {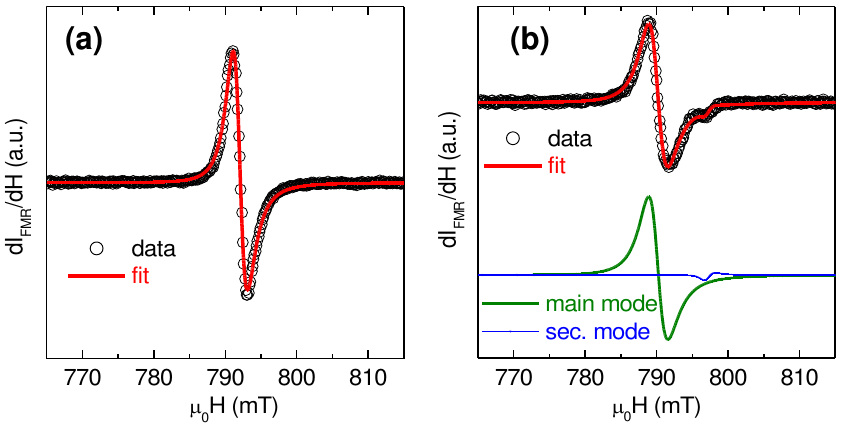}
  \centering
  \caption{\label{fig:lineshapeOP}
    Exemplary FMR spectra and fitting curves: (a) one mode of Lorentzian derivative; (b) superposition of a main mode and a small secondary mode due to slight sample inhomogeneity.}
\end{figure}

\begin{figure*}[t]
  \begin{minipage}[c]{0.66\textwidth}
    \includegraphics[width=\textwidth]{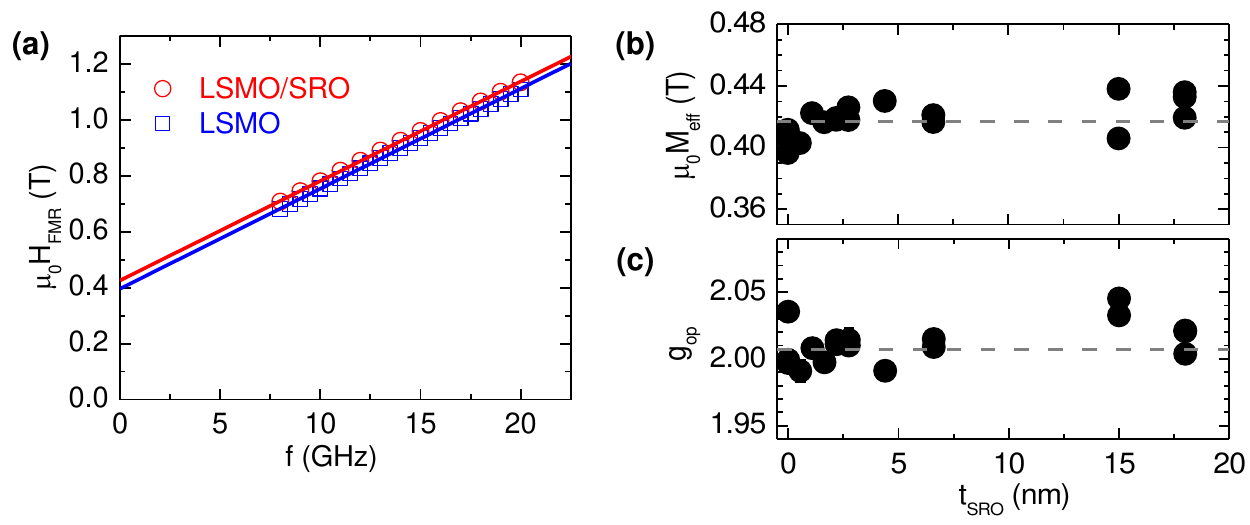}
  \end{minipage}\hfill
  \begin{minipage}[c]{0.33\textwidth}
   \caption{\label{fig:HfmrOP}
    (a) Out-of-plane resonance field $\HFMR$ versus excitation frequency $f$ for a single-layer LSMO(10 nm) film and a LSMO(10 nm)/SRO(3 nm) bilayer. The solid lines indicate fits to the data using Eq.~\ref{eq:KittelOP}. 
    (b,c) SRO-thickness dependence of the out-of-plane Land\'e g-factor (b) and effective saturation magnetization $\Meff$ (c). The dashed lines indicate the values averaged over all the data shown.}
  \end{minipage}
\end{figure*}
Broadband FMR measurements were performed at room temperature.
The film sample was placed face-down on a coplanar waveguide with a center conductor width of 250 $\mu$m. 
Each FMR spectrum was acquired at a constant excitation frequency while sweeping the external magnetic field $H$. 
The field derivative of the FMR absorption intensity (e.g., Fig.~\ref{fig:lineshapeOP}) was acquired using an rf diode combined with an ac (700 Hz) modulation field. 
Each FMR spectrum was fitted with the derivative of the sum of the symmetric and antisymmetric Lorentzians, as shown in Fig.~\ref{fig:lineshapeOP}, from which the resonance field $\HFMR$ and half-width-at-half-maximum linewidth $\Delta H$ were extracted. 
In some spectra (e.g., Fig.~\ref{fig:lineshapeOP}(b)), a small secondary mode in addition to the main FMR mode was observed. 
We fit such a spectrum to a superposition of two modes, each represented by a generalized Lorentzian derivative, and analyze only the $\HFMR$ and $\Delta H$ of the larger-amplitude main FMR mode. 
The secondary mode is not a standing spin-wave mode because it appears above or below the resonance field of the main mode $\HFMR$ with no systematic trend in field spacing. 
We attribute the secondary mode to regions in the film with slightly different \Ms or magnetic anisotropy. 
More pronounced inhomogeneity-induced secondary FMR modes have been observed in epitaxial magnetic films in prior reports~\cite{Haertinger2015, Luo2015}.

\section{Out-of-Plane FMR and Estimation of Spin Transport Parameters}\label{sec:OP}

\textcolor{black}{Out-of-plane FMR allows for conceptually simpler extraction of the static and dynamic magnetic properties of a thin-film sample. 
For fitting the frequency dependence of \HFMR, the Land\'e g-factor \gOP and effective saturation magnetization \Meff are the only adjustable parameters in the out-of-plane Kittel equation. 
The frequency dependence of \DelH for out-of-plane FMR arises solely from Gilbert damping, so that the conventional model of spin pumping~\cite{Tserkovnyak2002, Tserkovnyak2002a, Boone2013, Boone2015} can be used to analyze the data without complications from non-Gilbert damping. 
This consideration is particularly important because the linewidths of our LSMO(/SRO) films in in-plane FMR measurements are dominated by highly anisotropic non-Gilbert damping (as shown in Sec.~\ref{sec:IP}). 
Furthermore, a simple one-dimensional, time-independent model of spin pumping outlined by Boone \textit{et al.}~\cite{Boone2013} is applicable in the out-of-plane configuration, since the precessional orbit of the magnetization is circular to a good approximation. 
This is in contrast with the in-plane configuration with a highly elliptical orbit from a large shape anisotropy field.
By taking advantage of the simplicity in out-of-plane FMR, we find that the Gilbert damping parameter in LSMO is approximately doubled with the addition of a sufficiently thick SRO overlayer due to spin pumping. 
Our results indicate that spin-current transmission at the LSMO/SRO interface is comparable to previously reported ferromagnet/conductor bilayers and that spin diffusion length in SRO is $\lesssim$1 nm.}

We first quantify the static magnetic properties of LSMO(/SRO) from the frequency dependence of \HFMR. 
The Kittel equation for FMR in the out-of-plane configuration takes a simple linear form,
\begin{equation}\label{eq:KittelOP}
f=\frac{\gOP \muB}{h}\mu_0\left(\HFMR-\Meff\right), 
 \end{equation}
where $\mu_0$ is the permeability of free space, $\muB$ is the Bohr magneton, and $h$ is the Planck constant. 
As shown in Fig.~\ref{fig:HfmrOP}(a), we only fit data points where $\mu_0\HFMR$ is at least 0.2 T above $\mu_0 \Meff$ to ensure that the film is saturated out-of-plane. 
Figures~\ref{fig:HfmrOP}(b) and (c) plot the extracted $\Meff$ and \gOP, respectively, each exhibiting no significant dependence on SRO thickness $\tSRO$ to within experimental uncertainty. 
The SRO overlayer therefore evidently does not modify the bulk magnetic properties of LSMO, and significant interdiffusion across the SRO/LSMO interface can be ruled out. 
The averaged $\Meff$ of $330\pm10$ kA/m ($\mu_0 \Meff$ = $0.42\pm0.01$ T) is close to \Ms obtained from static magnetometery and implies negligible out-of-plane magnetic anisotropy; we thus assume $\Ms = \Meff$ in all subsequent analyses.  
The SRO-thickness independence of $\gOP$, averaging to $2.01\pm0.01$, implies that the SRO overlayer does not generate a significant orbital contribution to magnetism in LSMO. 
Moreover, the absence of detectable change in $\gOP$ with increasing \tSRO may indicate that the imaginary component of the spin-mixing conductance~\cite{Tserkovnyak2002a, Tserkovnyak2002} is negligible at the LSMO/SRO interface. 

\begin{figure*}[tb]
  \includegraphics [width=1.90\columnwidth] {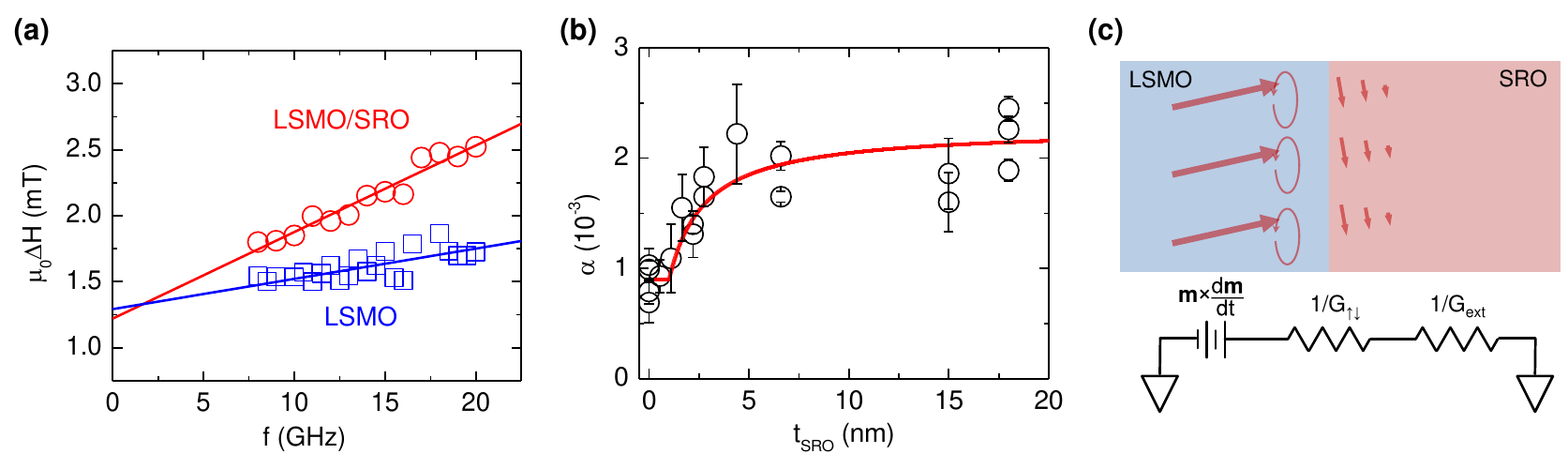}
  \centering
  \caption{\label{fig:linewidthOP}
    (a) Out-of-plane FMR linewidth $\Delta H$ versus excitation frequency for LSMO(10 nm) and LSMO(10 nm)/SRO(3 nm). The solid lines indicate fits to the data using Eq.~\ref{eq:Gilbert}. 
    (b) Gilbert damping parameter $\alpha$ versus SRO thickness \tSRO. The solid curve shows a fit to the diffusive spin pumping model (Eq.~\ref{eq:deltaA}).
(c) Schematic of out-of-plane spin pumping and the equivalent ``spin circuit.'' } 
\end{figure*}

The Gilbert damping parameter $\alpha$ is extracted from the frequency dependence of $\Delta H$ (e.g., Figure~\ref{fig:linewidthOP}(a)) by fitting the data with the standard linear relation,
\begin{equation}\label{eq:Gilbert}
\Delta H = \Delta H_{0} + \frac{h}{\gOP\muB}\alpha f. 
 \end{equation}
The zero-frequency linewidth $\Delta H_0$ is typically attributed to sample inhomogeneity.
We observe sample-to-sample variation of $\mu_0 \Delta H_0$ in the range $\approx1-4$ mT with no systematic correlation with $\tSRO$ or the slope in Eq.~\ref{eq:Gilbert}.  
Moreover, similar to the analysis of $\HFMR$, we only fit data obtained at $\geq$0.2 T above $\mu_0 \Meff$ to minimize spurious broadening of $\Delta H$ at low fields. 
The linear slope of $\Delta H$ plotted against frequency up to 20 GHz is therefore a reliable measure of $\alpha$ decoupled from $\Delta H_0$ in Eq.~\ref{eq:Gilbert}. 

Figure~\ref{fig:linewidthOP}(a) shows an LSMO single-layer film and an LSMO/SRO bilayer with similar $\Delta H_0$. 
The slope, which is proportional to $\alpha$, is approximately a factor of 2 greater for LSMO/SRO. 
Figure~\ref{fig:linewidthOP}(b) summarizes the dependence of $\alpha$ on SRO-thickness, \tSRO. 
For LSMO single-layer films we find $\alpha = (0.9\pm0.2)$$\times$$10^{-3}$, which is on the same order as previous reports of LSMO thin films~\cite{Luo2013, Luo2015, Lee2016, Wahler2016}.
This low damping is also comparable to the values reported in Heusler alloy thin films~\cite{Mizukami2009, Durrenfeld2015a} and may arise from the half-metal-like band structure of LSMO (Ref.~\citen{Liu2009a}). 
LSMO can thus be an efficient source of spin current generated resonantly by microwave excitation.
 
With a few-nanometer thick overlayer of SRO, $\alpha$ increases to $\approx$2$\times$10$^{-3}$ (Fig.~\ref{fig:linewidthOP}(b)). 
This enhanced damping with the addition of SRO overlayer may arise from (1) spin scattering~\cite{Nguyen2014, Rojas-Sanchez2014} at the LSMO/SRO interface or (2) spin pumping~\cite{Tserkovnyak2002, Tserkovnyak2002a} where nonequilibrium spins from LSMO are absorbed in the bulk of the SRO layer. 
Here, we assume that interfacial spin scattering is negligible, since $\lesssim$1 nm of SRO overlayer does not enhance $\alpha$ significantly (Fig.~\ref{fig:linewidthOP}(b)).
This is in contrast with the pronounced interfacial effect in ferromagnet/Pt bilayers~\cite{Nguyen2014, Rojas-Sanchez2014}, in which even $<$1 nm of Pt can increase $\alpha$ by as much as a factor of $\approx$2 (Refs.~\citen{Azzawi2016, Caminale2016, Emori2016}). 
In the following analysis and discussion, we show that spin pumping alone is sufficient for explaining the enhanced damping in LSMO with an SRO overlayer. 

We now analyze the data in Fig.~\ref{fig:linewidthOP}(b) using a one-dimensional model of spin pumping based on diffusive spin transport~\cite{Boone2013, Boone2015}.
The resonantly-excited magnetization precession in LSMO generates non-equilibrium spins polarized along  $\hat{m}\times \textrm{d}\hat{m}/\textrm{dt}$, which is transverse to the magnetization unit vector $\hat{m}$. 
This non-equilibrium spin accumulation diffuses out to the adjacent SRO layer and depolarizes exponentially on the characteristic length scale \sdl. 
The spin current density $\vec{j}_s$ at the LSMO/SRO interface can be written as~\cite{Polianski2004, Boone2013}
\begin{equation}\label{eq:spincurrent}
\vec{j}_s|_\text{interface} = \frac{\hbar^2}{2e^2}\frac{\hat{m}\times\frac{d\hat{m}}{dt}}{\left(\frac{1}{\Gmix}+\frac{1}{\Gext}\right)},
\end{equation}
where $\hbar$ is the reduced Planck constant, \Gmix is the interfacial spin-mixing conductance per unit area, and \Gext is the spin conductance per unit area in the bulk of SRO. 
In Eq.~\ref{eq:spincurrent}, 1/\Gmix and 1/\Gext constitute spin resistors in series such that the spin transport from LSMO to SRO can be regarded analogously as a ``spin circuit,'' as illustrated in Fig.~\ref{fig:linewidthOP}(c). 
In literature, these interfacial and bulk spin conductances are sometimes lumped together as an ``effective spin-mixing conductance'' $\Gmix^\mathrm{eff}=(1/\Gmix+1/\Gext)^{-1}$ (Refs.~\citen{Czeschka2011, Sun2013, Zhang2015a, Du2015,Jamali2015, Haertinger2015,Wahler2016,Lee2016,Lesne2016}). 
We also note that the alternative form of the (effective) spin-mixing conductance $g^\mathrm{(eff)}_\mathrm{eff}$, with units of m$^{-2}$, is related to $\Gmix^\mathrm{(eff)}$, with units of $\Omega^{-1}$m$^{-2}$, by $g^\mathrm{(eff)}_\mathrm{eff} = (h/e^2)\Gmix^\mathrm{(eff)}\approx 26$~k$\Omega \times \Gmix^\mathrm{(eff)}$. 

The functional form of \Gext is obtained by solving the spin diffusion equation with appropriate boundary conditions~\cite{Polianski2004, Boone2013, Boone2015}. 
In the case of a ferromagnet/nonmagnetic-metal bilayer, we obtain
\begin{equation}\label{eq:Gext}
\Gext = \frac{1}{2\rSRO\sdl}\tanh{\left(\frac{\tSRO}{\sdl}\right)},
\end{equation}
where \rSRO  is the resistivity of SRO, \tSRO is the thickness of the SRO layer, and \sdl is the diffusion length of pumped spins in SRO. 
Finally, the outflow of spin current (Eq.~\ref{eq:spincurrent}) is equivalent to an enhancement of Gilbert damping~\cite{Tserkovnyak2002} with respect to $\alpha_0$ of LSMO with $\tSRO=0$ such that
\begin{equation}\label{eq:deltaA}
\alpha = \alpha_0+\frac{\gOP\muB\hbar}{2e^2\Ms \tLSMO}{\left[\frac{1}{\Gmix}+{2\rSRO\sdl}\coth\left(\frac{\tSRO}{\sdl}\right)\right]}^{-1}.
\end{equation}
Thus, two essential parameters governing spin transport \Gmix and \sdl can be estimated by fitting the SRO-thickness dependence of $\alpha$ (Fig.~\ref{fig:linewidthOP}(b)) with Eq.~\ref{eq:deltaA}. 

In carrying out the fit, we fix $\alpha_0 = 0.9\times10^{-3}$.
We note that \rSRO increases by an order of magnitude compared to the bulk value of $\approx$2$\times$10$^{-6}$ $\Omega$m as \tSRO is reduced to a few nm; also, at thicknesses of 3 monolayers ($\approx$1.2 nm) or below, SRO is known to be insulating~\cite{Xia2009}.
We therefore use the \tSRO-dependent \rSRO shown in Appendix A while assuming \sdl is constant. 
An alternative fitting model that assumes a constant \rSRO, which is a common approach in literature, is discussed in Appendix A.

\begin{figure*}[tb]
  \includegraphics [width=1.90\columnwidth] {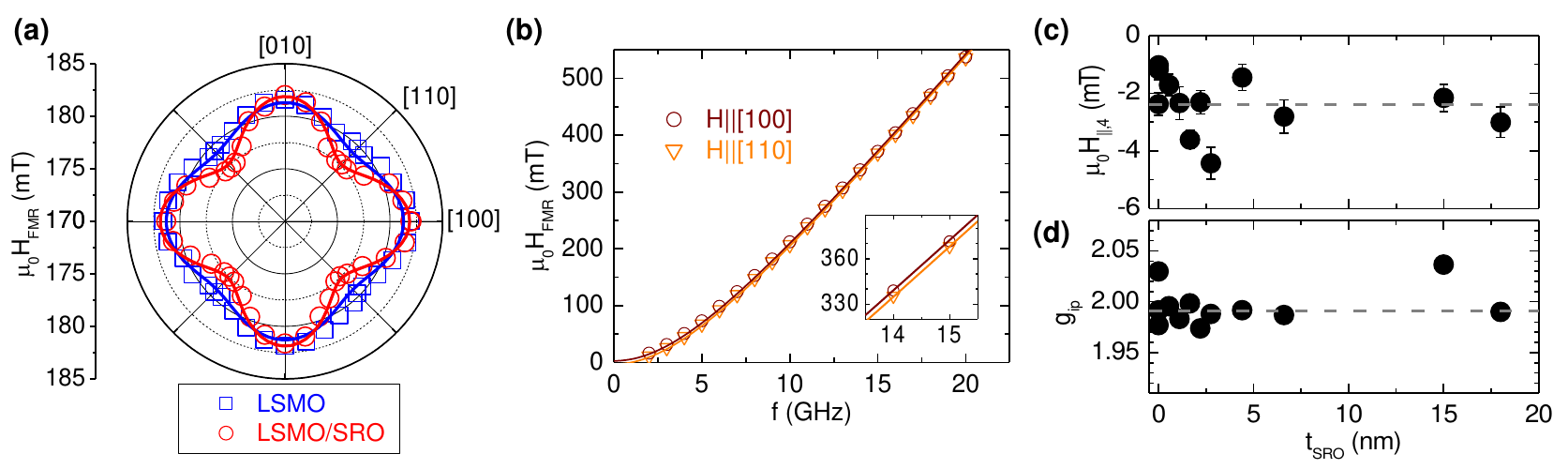}
  \centering
  \caption{\label{fig:HfmrIP}
    (a) Angular dependence of \HFMR at 9 GHz for LSMO(10 nm) and LSMO(10 nm)/SRO(7 nm). The solid curves indicate fits to the data using Eq.~\ref{eq:KittelIP}.
    (b) Frequency dependence of \HFMR for LSMO(10 nm)/SRO(7 nm) with field applied in the film plane along the [100] and [110] directions. Inset: close-up of \HFMR versus frequency around 14-15 GHz. In (a) and (b), the solid curves show fits to the Kittel equation (Eq.~\ref{eq:KittelIP}). 
    (c,d) SRO-thickness dependence of the in-plane cubic magnetocrystalline anisotropy field (c) and in-plane Land\'e g-factor (d). The dashed lines indicate the values averaged over all the data shown.} 
\end{figure*}

The curve in Fig.~\ref{fig:linewidthOP}(b) is generated by Eq.~\ref{eq:deltaA} with $\Gmix = 1.6\times10^{14}$ $\Omega^{-1}$m$^{-2}$  and $\sdl = 0.5$ nm.
Given the scatter of the experimental data, acceptable fits are obtained with $\Gmix \approx (1.2-2.5)\times10^{14}$ $\Omega^{-1}$m$^{-2}$ and $\sdl \approx 0.3-0.9$ nm.
The estimated ranges of \Gmix and \sdl also depend strongly on the assumptions behind the fitting model. 
For example, as shown in Appendix A, the constant-\rSRO model yields $\Gmix \gtrsim 3\times10^{14}$ $\Omega^{-1}$m$^{-2}$ and $\sdl \approx 2.5$ nm.
 
Nevertheless, we find that the estimated \Gmix is on the same order of magnitude as those of various ferromagnet/transition-metal heterostructures~\cite{Boone2015, Pai2015a, Montoya2016a}, signifying that the LSMO/SRO interface is reasonably transparent to spin current.
More importantly, the short \sdl implies the presence of strong spin-orbit coupling that causes rapid spin scattering within SRO. 
This finding is consistent with a previous study on SRO at low temperature in the ferromagnetic state showing extremely fast spin relaxation with Gilbert damping $\alpha\sim 1$ (Ref.~\citen{Langner2009}). 
The short \sdl indicates that SRO may be suitable as a spin sink or detector in all-oxide spintronic devices.

\section{In-plane FMR and Anisotropic Two-Magnon Scattering}\label{sec:IP}

In epitaxial thin films, the analysis of in-plane FMR is generally more complicated than that of out-of-plane FMR. 
High crystallinity of the film gives rise to a nonnegligible in-plane magnetocrystalline anisotropy field, which manifests in an in-plane angular dependence of \HFMR and introduces another adjustable parameter in the nonlinear Kittel equation for in-plane FMR. 
Moreover, \DelH in in-plane FMR of epitaxial thin films often depends strongly on the magnetization orientation and exhibits nonlinear scaling with respect to frequency due to two-magnon scattering, a non-Gilbert mechanism for damping~\cite{Lindner2003a, Woltersdorf2004a,Lenz2006, Zakeri2007a, Barsukov2011, Kurebayashi2013, Lee2016}. 
We indeed find that damping of LSMO in the in-plane configuration is anisotropic and dominated by two-magnon scattering.
We also observe evidence of enhanced two-magnon scattering with added SRO layers, which may be due to spin pumping from nonuniform magnetization precession.

Figure~\ref{fig:HfmrIP}(a) plots \HFMR of a single-layer LSMO film and an LSMO/SRO bilayer as a function of applied field angle within the film plane. 
For both samples, we observe clear four-fold symmetry, which is as expected based on the epitaxial growth of LSMO on the cubic LSAT(001) substrate. 
Similar to previous FMR studies of LSMO on SrTiO$_3$(001)~\cite{Belmeguenai2010, Monsen2014}, the magnetic hard axes (corresponding to the axes of higher \HFMR) are along $\langle$100$\rangle$. 
The in-plane Kittel equation for thin films with in-plane cubic magnetic anisotropy is~\cite{Farle1998},
\begin{equation}\label{eq:KittelIP}
\begin{split}
f=\frac{\gIP \mu_B}{h}\mu_0 &\left[\HFMR+\Hcub\cos(4\phi)\right]^{\frac{1}{2}}\times \\
&\left[\HFMR+\Meff+\dfrac{1}{4}\Hcub\left(3+\cos(4\phi)\right)\right]^{\frac{1}{2}},  
\end{split}
\end{equation}
where \gIP is the Land\'e g-factor that is obtained from in-plane FMR data,  \Hcub is the effective cubic anisotropy field, and $\phi$ is the in-plane field angle with respect to the [100] direction.
Given that LSMO is magnetically very soft (coercivity on the order of 0.1 mT) at room temperature, we assume that the magnetization is parallel to the field direction, particularly with $\mu_0H \gg 10$ mT. 
In fitting the angular dependence (e.g., Fig.~\ref{fig:HfmrIP}(a)) and frequency dependence (e.g., Fig.~\ref{fig:HfmrIP}(b)) of \HFMR to Eq.~\ref{eq:KittelIP}, we fix \Meff at the values obtained from out-of-plane FMR (Fig.~\ref{fig:HfmrOP}(b)) so that \Hcub and \gIP are the only fitting parameters. 
For the two samples shown in Fig.~\ref{fig:HfmrIP}(a), the fits to the angular dependence and frequency dependence data yield consistent values of \Hcub and \gIP. 
For the rest of the LSMO(/SRO) samples, we use the frequency dependence data with $H||[100]$ and $H||[110]$ to extract these parameters. 
Figures~\ref{fig:HfmrIP}(c) and (d) show that \Hcub and \gIP, respectively, exhibit no systematic dependence on \tSRO, similar to the findings from out-of-plane FMR (Figs.~\ref{fig:HfmrOP}(b),(c)).  
The in-plane cubic magnetocrystalline anisotropy in LSMO(/SRO) is relatively small, with $\mu_0$\Hcub averaging to $\approx$2.5 mT. 
\gIP averages out to $1.99\pm0.02$, which is consistent with \gOP found from out-of-plane FMR. 

\begin{figure*}
  \begin{minipage}[c]{0.63\textwidth}
    \includegraphics[width=\textwidth]{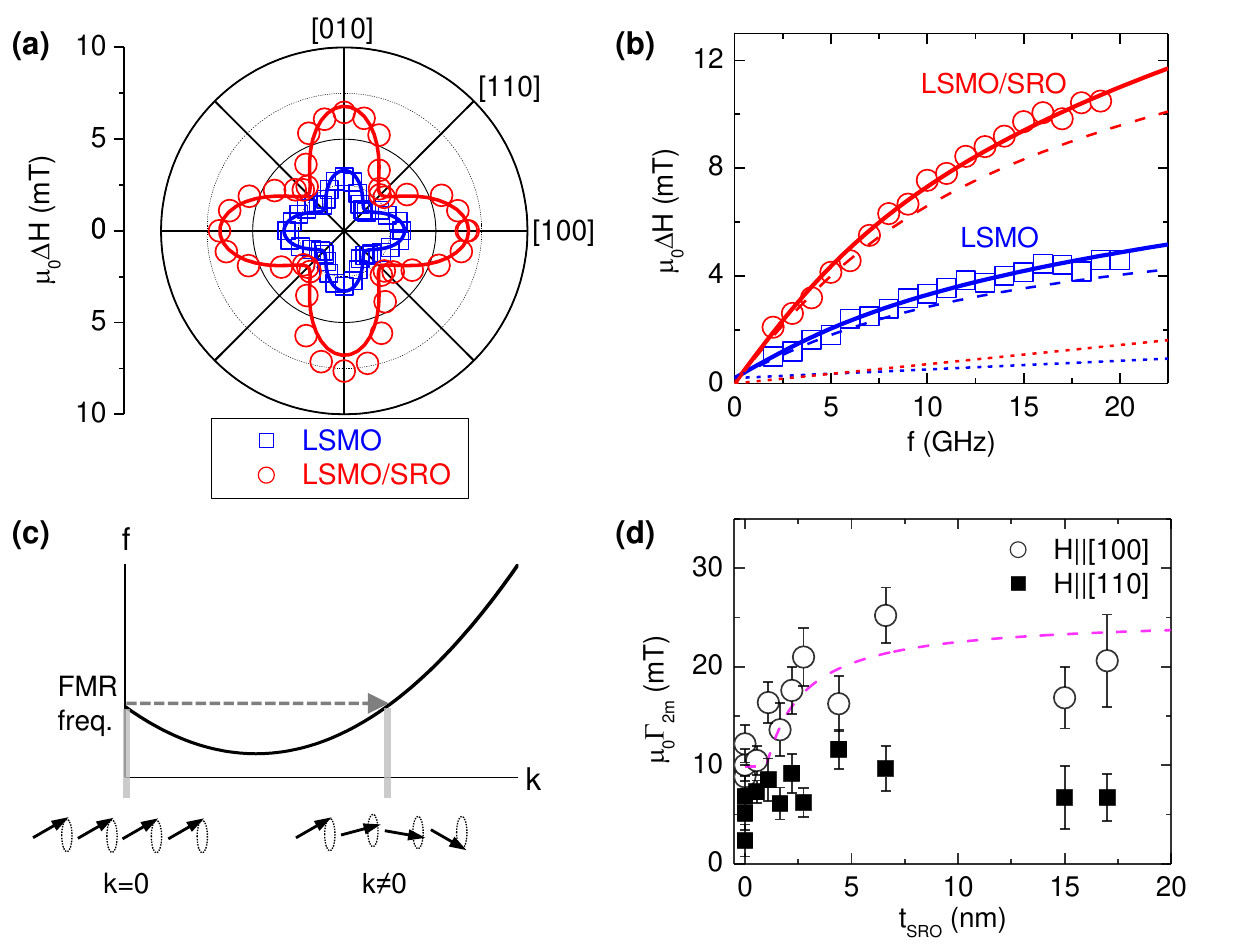}
  \end{minipage}\hfill
  \begin{minipage}[c]{0.37\textwidth}
   \caption{\label{fig:2mag}
    (a) In-plane angular dependence of linewidth $\Delta$H at 9 GHz for LSMO(10 nm) and LSMO(10 nm)/SRO(7 nm). The solid curves indicate fits to Eq.~\ref{eq:LWIPang}.
    (b) Frequency dependence of $\Delta$H for LSMO(10 nm) and LSMO(10 nm)/SRO(7 nm) with $H$ applied along the [100] direction. The solid curves indicate fits to Eq.~\ref{eq:2mag}. The dashed and dotted curves indicate estimated two-magnon and Gilbert damping contributions, respectively.
    (c) Schematic of a spin wave dispersion curve (when the magnetization is in-plane and has a finite component parallel to the spin wave wavevector $k$) and two-magnon scattering.
    (d) Two-magnon scattering coefficient \Gtm, estimated for the cases with $H$ applied along the [100] and [110] axes, plotted against SRO thickness \tSRO. The dashed curve is the same as that in Fig.~\ref{fig:linewidthOP}(c) scaled to serve as a guide for the eye for \Gtm with H along [100].}
  \end{minipage}
\end{figure*}

While the magnetocrytalline anisotropy in LSMO(/SRO) is found to be modest and independent of \tSRO, we observe much more pronounced in-plane anisotropy and \tSRO dependence in linewidth \DelH, as shown in Figs.~\ref{fig:2mag}(a) and (b). 
Figure~\ref{fig:2mag}(a) indicates that the in-plane dependence of \DelH is four-fold symmetric for both LSMO(10 nm) and LSMO(10 nm)/SRO(7 nm). 
\DelH is approximately a factor of 2 larger when the sample is magnetized along $\langle$100$\rangle$ compared to when it is magnetized along  $\langle$110$\rangle$.
One might attribute this pronounced anisotropy to anisotropic Gilbert damping~\cite{Baker2016}, such that the sample magnetized along the hard axes $\langle$100$\rangle$ may lead to stronger damping. 
However, we find no general correlation between magnetocrystalline anisotropy and anisotropic \DelH: As we show in Appendix B, LSMO grown on NdGaO$_3$(110) with pronounced uniaxial magnetocrystalline anisotropy exhibits identical \DelH when magnetized along the easy and hard axes. 
Moreover, whereas Gilbert damping should lead to a linear frequency dependence of \DelH, for LSMO(/SRO) the observed frequency dependence of \DelH is clearly nonlinear as evidenced in Fig.~\ref{fig:2mag}(b). 
The pronounced anisotropy and nonlinear frequency dependence of \DelH together suggest the presence of a different damping mechanism. 

A well-known non-Gilbert damping mechanism in highly crystalline ultrathin magnetic films is two-magnon scattering~\cite{Mills2002,  Arias1999, McMichael1998, Lindner2003a, Woltersdorf2004a,Lenz2006, Zakeri2007a, Barsukov2011, Kurebayashi2013, Lee2016}, in which uniformly precessing magnetic moments (a spin wave, or magnon mode, with wavevector $k = 0$) dephase to a $k \neq 0$ magnon mode with adjacent moments precessing with a finite phase difference. 
By considering both exchange coupling (which results in magnon energy proportional to $k^2$) and dipolar coupling (magnon energy proportional to $-$$|k|$) among precessing magnetic moments, the $k=0$ and $k\neq0$ modes become degenerate in the magnon dispersion relation~\cite{Mills2002} as illustrated in Fig.~\ref{fig:2mag}(c).  

The transition from $k=0$ to $k\neq0$ is activated by defects that break the translational symmetry of the magnetic system by localized dipolar fields~\cite{Mills2002, Arias1999, McMichael1998}. 
In LSMO(/SRO), the activating defects may be faceted such that two-magnon scattering is more pronounced when the magnetization is oriented along $\langle$100$\rangle$. 
One possibility is that LSMO thin films naturally form pits or islands faceted along $\langle$100$\rangle$ during growth. 
However, we are unable to consistently observe signs of such faceted defects in LSMO(/SRO) samples with an atomic force microscope (AFM).
It is possible that these crystalline defects are smaller than the lateral resolution of our AFM setup ($\lesssim$10 nm) or that these defects are not manifested in surface topography. 
Such defects may be point defects or nanoscale clusters of distinct phases that are known to exist intrinsically even in high-quality crystals of LSMO (Ref.~\citen{Dagotto2003}).

Although the definitive identification of defects that drive two-magnon scattering would require further investigation, we can rule out (1) atomic step terraces and (2) misfit dislocations as sources of anisotropic two-magnon scattering. 
(1) AFM shows that the orientation and density of atomic step terraces differ randomly from sample to sample, whereas the anisotropy in \DelH is consistently cubic with larger \DelH for $H||\langle$100$\rangle$ than $H||\langle$110$\rangle$. 
This is in agreement with the recent study by Lee \textit{et al.}, which shows anisotropic two-magnon scattering in LSMO to be independent of regularly-spaced parallel step terraces on a buffered-oxide etched SrTiO$_3$ substrate~\cite{Lee2016}.
(2) Although Woltersdorf and Heinrich have found that misfit dislocations in Fe/Pd grown on GaAs are responsible for two-magnon scattering~\cite{Woltersdorf2004a}, such dislocations are expected to be virtually nonexistent in fully strained LSMO(/SRO) films on the closely-latticed matched LSAT substrates~\cite{Takamura2008, Grutter2010}. 

We assume that the in-plane four-fold anisotropy and nonlinear frequency dependence of \DelH are entirely due to two-magnon scattering.
For a sample magnetized along a given in-plane crystallographic axis $\langle$hk0$\rangle$ = $\langle$100$\rangle$ or $\langle$110$\rangle$, the two-magnon scattering contribution to \DelH is given by~\cite{Arias1999}
\begin{equation}\label{eq:2mag}
\DelHtmhkl = \Gtm^{\langle \mathrm{hk0} \rangle}\sin^{-1}\sqrt{\frac{\sqrt{f^2+(\fM/2)^2}-\fM/2}{\sqrt{f^2+(\fM/2)^2}+\fM/2}},
\end{equation}
where $\fM = (\gIP\muB/h)\mu_0\Ms$ and $\Gtm^{\langle\mathrm{hk0}\rangle}$ is the two-magnon scattering parameter. 
The angular dependence of \DelH is fitted with~\cite{Woltersdorf2004a}
\begin{equation}\label{eq:LWIPang}
\begin{split}
\DelH&=\DelH_0+\frac{h}{\gIP\muB}\alpha f\\
&+\DelHtm^{\langle100\rangle}\cos^2(2\phi)+\DelHtm^{\langle110\rangle}\cos^2(2[\phi-\tfrac{\pi}{4}]).
\end{split}
\end{equation}
Similarly, the frequency dependence of \DelH with the sample magnetized along [100] or [110], i.e., $\phi=0$ or $\pi/4$, is well described by Eqs.~\ref{eq:2mag} and \ref{eq:LWIPang}.
In principle, it should be possible to fit the linewidth data with \DelHo, $\alpha$, and \Gtm as adjustable parameters. 
In practice, the fit carried out this way is overspecified such that wide ranges of these parameters appear to fit the data. 
We therefore impose a constraint on $\alpha$ by assuming that Gilbert damping for LSMO(/SRO) is isotropic: For each SRO thickness \tSRO, $\alpha$ is fixed to the value estimated from the fit curve in Fig.~\ref{fig:linewidthOP}(c) showing out-of-plane FMR data. 
(This assumption is likely justified, since the damping for LSMO(10 nm) on NdGdO$_3$(110) with strong uniaxial magnetic anisotropy is identical for the easy and hard directions, as shown in Appendix B.)
To account for the uncertainty in the Gilbert damping in Fig.~\ref{fig:linewidthOP}(c), we vary $\alpha$ by $\pm$25\% for fitting the frequency dependence of in-plane \DelH. 
Examples of fits using Eqs.~\ref{eq:2mag} and \ref{eq:LWIPang} are shown in Fig.~\ref{fig:2mag}(a),(b).

Figure~\ref{fig:2mag}(d) shows that the SRO overlayer enhances the two-magnon scattering parameter \Gtm by up to a factor of $\approx$2 for $H||$[100]. 
By contrast, for $H||$[110], although LSMO/SRO exhibits enhanced \DelH compared to LSMO, the enhancement in \Gtm is obscured by the uncertainty in Gilbert damping.
In Table I, we summarize the Gilbert and two-magnon contributions to \DelH for LSMO single layers and LSMO/SRO (averaged values for samples with $\tSRO > 4$ nm) with $H||$[100] and $H||$[110]. 
Comparing the effective spin relaxation rates, $({\gIP\muB}/{h})\mu_0\Ms\alpha$ and $({\gIP\muB}/{h})\mu_0\Gtm$, reveals that two-magnon scattering dominates over Gilbert damping.  

\begin{center}
\begin{table}[b]
\begin{threeparttable}
\caption{Spin relaxation rates extracted from in-plane FMR ($10^6$ s$^{-1}$)} 
\centering 
\begin{tabular}{c ccccc} 
\hline\hline 
{}&\multicolumn{2}{c}{LSMO}&\multicolumn{2}{c}{LSMO/SRO*} \\ [0.5ex] 
\hline 
\\
Gilbert: $\frac{\gIP\muB}{h}\mu_0\Ms\alpha$ & \multicolumn{2}{c}{$11\pm2$} & \multicolumn{2}{c}{$23\pm4$}\\
\\
two-magnon: $\frac{\gIP\muB}{h}\mu_0\Gtm$ $(H||[100])$ & \multicolumn{2}{c}{$290\pm50$} & \multicolumn{2}{c}{$550\pm100$}\\
\\
two-magnon: $\frac{\gIP\muB}{h}\mu_0\Gtm$ $(H||[110])$ & \multicolumn{2}{c}{$140\pm60$} & \multicolumn{2}{c}{$250\pm60$}\\[1ex] 
\hline 
\end{tabular}
\begin{tablenotes}
      \item * Averaged over samples with $\tSRO>4$ nm. 
\end{tablenotes}
\end{threeparttable}
\label{tab:params}
\end{table}
\end{center}

We now speculate on the mechanisms behind the enhancement in \Gtm in LSMO/SRO, particularly for $H||[100]$. 
One possibility is that SRO interfaced with LSMO directly increases the rate of two-magnon scattering, perhaps due to formation of additional defects at the surface of LSMO. 
If this were the case we might expect a significant increase and saturation of \Gtm at small \tSRO.
However, in reality, \Gtm increases for $\tSRO > 1$ nm (Fig.~\ref{fig:2mag}(d)), which suggests spin scattering in the bulk of SRO. 
We thus speculate another mechanism, where $k\neq 0$ magnons in LSMO are scattered by spin pumping into SRO. 
As shown by the guide-for-the-eye curve in Fig.~\ref{fig:2mag}(d), the \tSRO dependence of \Gtm (for $H||$[100]) may be qualitatively similar to the \tSRO dependence of $\alpha$ measured from out-of-plane FMR (Fig.~\ref{fig:linewidthOP}(c)); this correspondence would imply that the same spin pumping mechanism, which is conventionally modeled to act on the $k = 0$ mode, is also operative in the degenerate $k \neq 0$ magnon mode in epitaxial LSMO. 
Indeed, previous studies have electrically detected the presence of spin pumping from $k \neq 0$ magnons by the inverse spin-Hall effect in Y$_3$Fe$_5$O$_{12}$/Pt bilayers~\cite{Sandweg2011, DaSilva2013, Manuilov2015}.
However, we cannot conclusively attribute the observed FMR linewidth broadening in LSMO/SRO to such $k \neq 0$ spin pumping, since it is unclear whether faster relaxation of $k \neq 0$ magnons should necessarily cause faster relaxation of the $k=0$ FMR mode. 
Regardless of its origin, the pronounced anisotropic two-magnon scattering introduces additional complexity to the analysis of damping in LSMO/SRO and possibly in other similar ultrathin epitaxial magnetic heterostructures.

\section{Summary}
We have demonstrated all-oxide perovskite bilayers of LSMO/SRO that form spin-source/spin-sink systems.
From out-of-plane FMR, we deduce a low Gilbert damping parameter of $\approx$1$\times$10$^{-3}$ for LSMO.
The two-fold enhancement in Gilbert damping with an SRO overlayer is adequately described by the standard model of spin pumping based on diffusive spin transport. 
We arrive at an estimated spin-mixing conductance $\Gmix \approx (1-2)\times10^{14}$ $\Omega^{-1}$m$^{-2}$ and spin diffusion length $\sdl \lesssim 1$ nm, which indicate reasonable spin-current transparency at the LSMO/SRO interface and strong spin scattering within SRO. 
From in-plane FMR, we reveal pronounced non-Gilbert damping, attributed to two-magnon scattering, which results in a nonlinear frequency dependence and anisotropy in linewidth. 
The magnitude of two-magnon scattering increases with the addition of an SRO overlayer, pointing to the presence of spin pumping from nonuniform spin wave modes. 
Our findings lay the foundation for understanding spin transport and magnetization dynamics in epitaxial complex oxide heterostructures.

\section*{Acknowledgements}
We thank Di Yi, Sam Crossley, Adrian Schwartz, Hankyu Lee, and Igor Barsukov for helpful discussions, and Tianxiang Nan and Nian Sun for the design of the coplanar waveguide. This work was funded by the National Security Science and Engineering Faculty Fellowship of the Department of Defense under Contract No. N00014-15-1-0045.

\section*{Appendix A: Spin Pumping and SRO Resistivity}

\begin{figure}[tb]
  \includegraphics [width=1.00\columnwidth] {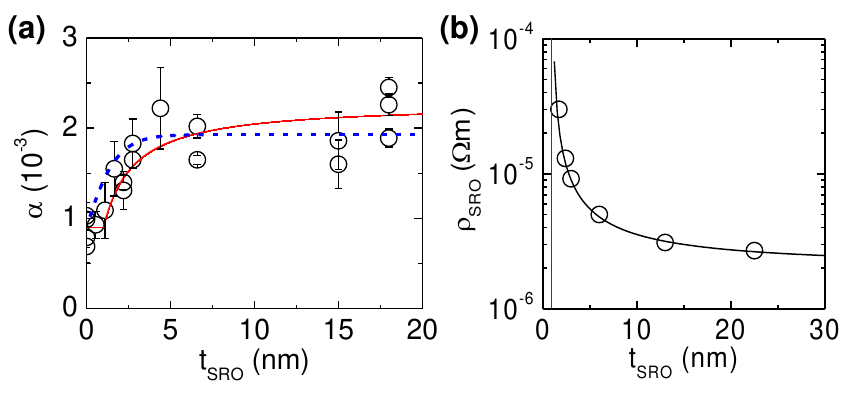}
  \centering
  \caption{\label{fig:altSP}
    (a) Gilbert damping parameter $\alpha$ versus SRO thickness \tSRO. The solid curve is a fit taking into account the \tSRO dependence of SRO resistivity, whereas the dotted curve is a fit assuming a constant bulk-like SRO resistivity. (b) Resistivity of SrRuO$_3$ films on LSAT(001) as a function of thickness. } 
\end{figure}

When fitting the dependence of the Gilbert damping parameter $\alpha$ on spin-sink thickness, a constant bulk resistivity for the spin sink layer is often assumed in literature. 
By setting the resistivity of SRO to the bulk value $\rSRO = 2\times10^{-6}\ \Omega$m and fitting the $\alpha$-versus-\tSRO data (Fig.~\ref{fig:linewidthOP}(c) and reproduced in Fig.~\ref{fig:altSP}(a)) to Eq.~\ref{eq:deltaA}, we arrive at $\Gmix \gtrsim  3\times10^{14}$ $\Omega^{-1}$m$^{-2}$ and $\sdl \approx 2.5$ nm.
The fit curve is insensitive to larger values of \Gmix because the bulk spin resistance 1/\Gext, with the relatively large resistivity of SRO, dominates over the interfacial spin resistance 1/\Gmix (see Eqs.~\ref{eq:Gext} and \ref{eq:deltaA}). 
As shown by the dotted curve in Fig.~\ref{fig:altSP}, this simple constant-\rSRO model appears to mostly capture the \tSRO-dependence of $\alpha$. 
This model of course indicates finite spin pumping at even very small SRO thickness $\lesssim$1 nm, which is likely nonphysical since SRO should be insulating in this thickness regime~\cite{Xia2009}. 
Indeed, \sdl estimated with this model should probably be considered a phenomenological parameter: As pointed out by recent studies, strictly speaking, a physically meaningful estimation of \sdl should take into account the thickness dependence of the resistivity of the spin sink layer~\cite{Boone2015, Nguyen2016, Montoya2016a}, especially for SRO whose thickness dependence of resistivity is quite pronounced.

\begin{figure}[tb]
  \includegraphics [width=1.00\columnwidth] {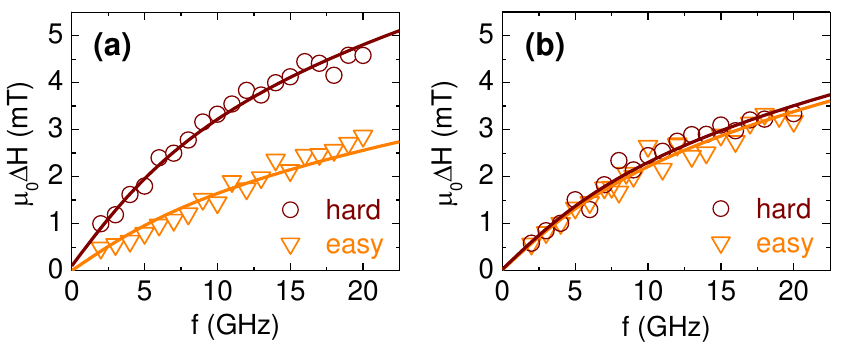}
  \centering
  \caption{\label{fig:LSMOsubs}
    Frequency dependence of in-plane FMR linewidth \DelH of LSMO(10 nm) on (a) LSAT(001) and (b) NdGaO$_3$(110), with the magnetization along the magnetic easy and hard axes. The solid curves are fits to Eq.~\ref{eq:2mag} with the Gilbert damping parameter $\alpha$ fixed to $0.9\times10^{-3}$.} 
\end{figure}

Figure~\ref{fig:altSP}(b) plots the SRO-thickness dependence of the resistivity of SRO films deposited on LSAT(001) measured in the four-point van der Pauw geometry.
The trend can be described empirically by
\begin{equation}\label{eq:resistivity}
\rSRO = \rbulk+\frac{\rsurf}{\tSRO-\tTh},
\end{equation}
where $\rbulk = 2\times10^{-6}\ \Omega$m is the resistivity of SRO in the bulk limit, $\rsurf = 1.4\times10^{-14}\ \Omega$m$^2$ is the surface resistivity coefficient, and $\tTh = 1$ nm is the threshold thickness below which the SRO layer is essentially insulating.   
The value of \tTh agrees with literature reporting that SRO is insulating at thickness of 3 monolayers ($\approx$1.2 nm) or below~\cite{Xia2009}. 
Given the large deviation of \rSRO from the bulk value, especially at small \tSRO, the trend in Fig.~\ref{fig:altSP}(b) suggests that taking into account the \tSRO dependence of \rSRO is a sensible approach.

\section*{Appendix B: In-Plane Damping of LSMO on Different Substrates}

In Fig.~\ref{fig:LSMOsubs}, we compare the frequency dependence of \DelH for 10-nm thick LSMO films deposited on different substrates: LSAT(001) and NdGaO$_3$(110). (NdGaO$_3$ is an orthorhombic crystal and has a $\sqrt{2}$-pseudocubic parameter of $\approx$3.86 \AA, such that (001)-oriented LSMO grows on the (110)-oriented surface of NdGaO$_3$.) 
As shown in Sec.~\ref{sec:IP}, LSMO on LSAT(001) exhibits cubic magnetic anisotropy within the film plane with the $\langle110\rangle$ and $\langle100\rangle$ as the easy and hard axes, respectively. 
LSMO on NdGaO$_3$(110) exhibits uniaxial magnetic anisotropy within the film plane with $[1\bar{1}0]$ and [001] of NdGaO$_3$ as the easy and hard axes, respectively.~\cite{Boschker2009}
Whereas LSMO on LSAT(001) shows distinct magnitudes of damping when the film is magnetized along the easy and hard axes (Figs.~\ref{fig:LSMOsubs}(a) and \ref{fig:2mag}(a)), in LSMO on NdGaO$_3$(110) damping is identical for both the easy and hard axes (Figs.~\ref{fig:LSMOsubs}(b)). 
These results demonstrate that higher damping (wider linewidth) is in general not linked to the magnetic hard axis of LSMO.

\end{document}